\documentclass[twocolumn,letter]{jpsj2} 

\def\x{{\bf x}}
\def\y{{\bf y}}
\def\z{{\bf z}}

\title{Testing Error Correcting Codes \\
by Multicanonical Sampling of Rare Events}
\author{Yukito Iba and Koji Hukushima}
\inst{
The Institute of Statistical Mathematics
4-6-7 Minami-Azabu, Minato-ku, Tokyo 106-8569, Japan \\
Department of Basic Science, University of Tokyo, 
3-8-1 Komaba, Meguro-ku, Tokyo 153-8902, Japan}

\abst{The idea of rare
event sampling by  Multicanonical Monte Carlo
is applied to the estimation of the 
performance of error-correcting codes. 
The essence of the idea is importance sampling of the pattern of noises 
in the channel by Multicanonical Monte Carlo, which enables
efficient estimation of tails of the distribution of bit errors.
The proposed method is successfully 
tested with a convolutional code.
}

\begin{document}
\maketitle

Dynamic Monte Carlo (Markov chain Monte
Carlo, {\sf MCMC}) algorithm is discovered in physics
in 1950s and introduced to statistical data analysis in 
1980s, now being recognized as an essential methodology 
in both fields. But the use of {\sf MCMC} need not
be restricted to these two fields. 
In fact, it is a general strategy for sampling from complicated
distributions with unknown normalization constants, and
there will be a number of potential applications in
other fields. 

In this note, we discuss
a problem of estimating the distribution of bit errors 
of error-correcting codes~\cite{Gbook,Mbook,Nbook} for a given
channel and a given input distribution. 
The essence of the proposed idea 
is {\it importance sampling of the pattern of noises 
in the channel by MCMC\/}. 
By using the proposed strategy, the tails of the error 
distribution are efficiently calculated.
We also discuss that the use of
Multicanonical Monte Carlo algorithm\cite{BergN,BergC,Lee,iba}, 
a version of MCMC
which utilizes iterative construction of the 
sampling weight,
is ideally suited to the problem.

Note that efficient generation of
unusual events is important not only for checking
theories but also for practical purposes, because
``rare events'' become no more rare with 
some deviation from idealized models, such as 
unexpected correlation between noises.  

Our proposal is on the use of 
MCMC for the evaluation of characteristics of
codes and channels. Thus, it is essentially 
different from any idea on the use of 
MCMC for decoding messages.
An example of the use
of MCMC and related algorithm in this field 
is found in the references\cite{cap1,cap2}, which estimate 
channel capacity by Blahut-Arimoto algorithm.
 
The proposed approach is closely related to the idea of
Hartmann~\cite{Hartmann2002}, who studied large deviations 
of the output of a sequence alignment algorithm by 
a MCMC sampling. Similar
approaches based on a MCMC sampling of quenched disorder
is also used for estimating tails of 
the distributions of the ground state energy of
physical systems~\cite{Keorner2006,Monthus2006} and
exploring finite temperature property of random magnets\cite{H07,N07}. 
The idea of using MCMC as a tool for sampling rare events 
that reduce the worst-case efficiency of an
algorithm seems to have a wide range of applications.

Let us introduce a problem of estimating the 
performance of error-correcting codes. 
Fix a coding, a channel, and a decoding algorithm,
and denote an input message and its distribution 
by $\x=\{x_i\}$ and $P(\x)$. The encoded message,
an output of the noisy channel and its distribution is 
represented as $\z=\{z_i\}$, $\y=\{y_i\}$ and 
$P(\y|\x)$, respectively. The message decoded by the given
method from $\y$ is given by $T(\y)$.
Here any kind of decoding algorithm 
is allowed such as Viterbi-decoding, belief-propagation, and
loopy-belief-propagation, as long as $T(\y)$ is stable and can be
regarded as a deterministic decoder. Assuming 
the distance $d(\x,T(\y))$ 
between the input $\x$ and output $T(\y)$, we consider the problem of
calculating the probability distribution 
\begin{equation} \label{defpd}
P(d)=\sum_{\x,\y}
\delta(d-d(\x,T(\y)))P(\x)P(\y|\x)
\end{equation}
of the distance $d$ between inputs and outputs. 
In most familiar
cases, $d$ is Hamming distance and $P(d)$ is the distribution
of bit errors with a given distribution $P(\x)$ of inputs.

A naive method for estimating $P(d)$
is repeating the following procedures (1--3)
independently until desired accuracy is attained:  
(1)~A message $\x$ is generated from the distribution $P(\x)$.
(2)~The output $\y$ of the channel is sampled from $P(\y|\x)$.
(3)~The distance $d(\x,T(\y))$ is calculated and recorded. 
This method is straightforward, but becomes highly computationally 
expensive when we are interested in the tails of the distribution 
of $P(d)$, which correspond to {\it rare events} or {\it large 
deviations} under the assumption of the 
distribution $P(\x)$ of inputs.

The proposed method, which largely improves the 
efficiency of the estimation
of the tails of $P(d)$, is multicanonical sampling~\cite{BergN,BergC,Lee,iba}
with a weight $w_{\rm M}(\x,\y)$ approximately proportional to  
$P(d(\x,T(\y)))^{-1} P(\x)P(\x|\y)$. Here $P(d(\x,T(\y)))$
is defined by the expression where $d$ of $P(d)$ 
is substituted for $d(\x,T(\y))$. 
From the definition (\ref{defpd})
of $P(d)$, it is shown that 
the marginal distribution $P^*(d)$ of bit errors $d$ 
with this weight becomes nearly flat on the interval on which
$P(d) \neq 0$, i.e.,
\begin{eqnarray*}
\lefteqn{P^*(d) \simeq} & \\  
 & {\rm c} \,\sum_{\x,\y}
\delta(d-d(\x,T(\y))) P(d(\x,T(\y)))^{-1} P(\x)P(\x|\y) \\
& =  {\rm c} \,\,
P(d)^{-1} \sum_{\x,\y} \delta(d-d(\x,T(\y))) P(\x)P(\x|\y) 
=  {\rm c} {}_,
\end{eqnarray*}
where $c$ is a constant.
This enables both efficient sampling of the tails of 
the distribution $P(d)$ and fast mixing of the 
Markov chain used for sampling. 
In this generic form, the method contains sampling of
both $\x$ and $\y$ and looks somewhat complicated.
But in the examples discussed below, it becomes
simpler and reduces to the original idea of sampling
pattern of the noise in the channel.

How can we realize such a sampling and fit
the results of the sampling 
into the original problem?
If we define
the function $P_{\rm M}(d)$ by
\begin{equation} \label{wm}
w_{\rm M}(\x,\y)=P_{\rm M}(d(\x,T(\y)))^{-1} P(\x)P(\x|\y) {}_,
\end{equation}
the choice of weight $w_{\rm M}(\x,\y)$
reduces to the estimation of $P_{\rm M}(d)$ that realizes
almost flat marginal distribution $P^{*}(d)$. Then $P_{\rm M}(d)$
can be estimated by repeated 
preliminary runs of the simulation, just in 
the same way as the estimation of the weight 
in a conventional multicanonical algorithm.
While any method in literature of multicanonical or Wang-Landau
algorithm can be used for the estimation of the weight, 
in the following example we use a naive method with a 
histogram construction (entropic sampling)~\cite{Lee,iba}.
Once we obtain $P_{\rm M}(d)$ that gives a sufficiently  
flat distribution $P^*(d)$, 
a reconstruction of the target distribution $P(d)$ is
given by $P^{*}(d) P_{\rm M}(d)$ with a suitable normalization
constant.

Here we test the proposed method with a convolutional
code, whose codewords are $z^{(1)}_i=x_ix_{i+2}$ and
$z^{(2)}_i=x_ix_{i+1}x_{i+2}$. A binary symmetric channel
(BSC) is assumed and a Viterbi decoder is used as
$T(\y)$. In this example, the gauge invariance~\cite{Nbook}
considerably simplifies the algorithm. In particular, we can
fix the input $x$ to an arbitrary bit sequence such as $\x_0=
1111111 \cdots$ and the sampling of $\x$ and $\y$ reduces 
to the sampling of the pattern $\y$ of noises
in the channel~\cite{xxx}. Then the expression (\ref{wm})
becomes
$$
w_{\rm M}(\y)=P_{\rm M}(d(\x_0,T(\y)))^{-1} P(\x_0|\y) {}_,
$$
but the proposed algorithm can be applied
with some obvious modifications.
In the following example,
the length of original message and 
encoded message is $200$ and $(200-2)\cdot 2=396$ respectively, and 
the probability of bit flip
is set to $0.1$. 20 iterations
of preliminary runs are performed 
for the tuning of the weight and the final run 
is used for the calculation of the results.

Fig.1 gives an example of the convergence of algorithm, where
the estimated bit error probabilities
after the $4$th, $12$th, and $20$th iteration are shown by solid circles 
\verb+(+$\bullet$\verb+)+,
triangles \verb+(+$\triangle$\verb+)+, 
and open circles \verb+(o)+, respectively. 
The horizontal part of each curve indicates that
no sample is obtained in the region.

In Fig.2, probabilities estimated in the $20$th iteration
are compared with the one by the naive method 
based on uniform random 
sampling. The symbol \verb+(+$\bullet$\verb+)+ corresponds to 
the result by the proposed method, where
total of the preliminary and measurement runs 
requires about $8 \times 10^8$ Viterbi
decoding. The symbol \verb|(+)| 
corresponds to the result by the naive method
with $8.08 \times 10^8$ Viterbi decoding. 
The results by the naive method are not shown 
in the region where the method dose not give 
a sample. 

The proposed method gives the result in Fig.2 
within a day of computation by a current personal computer and 
enables sampling from the right tail of the probability distribution 
$P(d)$ where the naive method can hardly realize. 
On the other hand, the result by the proposed method \verb+(+$\bullet$\verb+)+ 
agrees with the one by the naive method \verb|(+)| based on uniform random 
sampling in the range of higher probabilities, which supports
the validity of the proposed method.

\begin{figure}[htbp]
\vspace*{-0.3cm}
\begin{center}
\includegraphics[width=7cm]{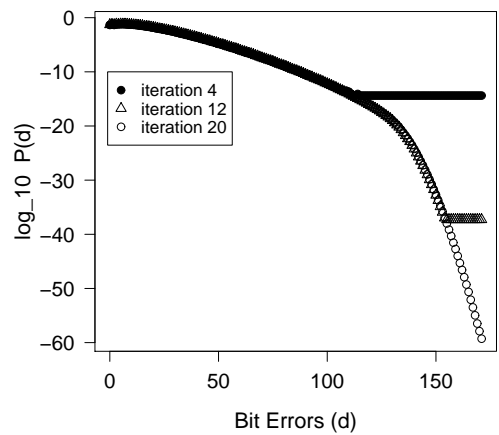}
\end{center}
\caption{An example of convergence of the proposed algorithm.}
\end{figure}

\begin{figure}[htbp]
\vspace*{-0.3cm}
\begin{center}
\includegraphics[width=7cm]{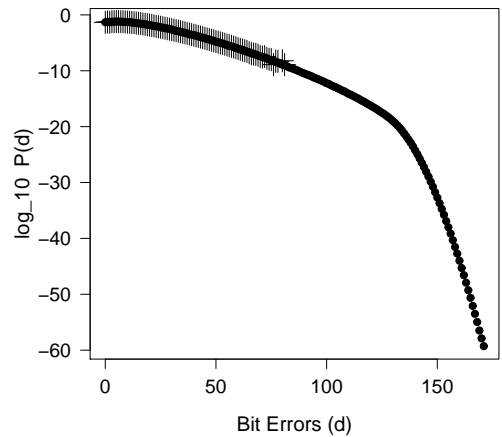}
\end{center}
\caption{Comparison of the estimates
of $P(d)$ calculated by the proposed method ($\bullet$)
and the naive method (+). Binary symmetric channel.}

\end{figure}

In the case of  binary symmetric channel, 
the probability of bit flips is
given and there is a small but finite possibility of
flipping arbitrary number of bits up to the length of the
encoded message.  
Then it is natural to expect that the right 
tail of the distribution $P(d)$ corresponds to larger number
of flipped bits in the channel. 
This tendency is illustrated in Fig.3, where the average
of the numbers of flipped bits conditioned with 
a given value of bit errors $d$
is plotted as a function of $d$.

\begin{figure}[htbp]
\vspace*{-0.3cm}
\begin{center}
\includegraphics[width=7cm]{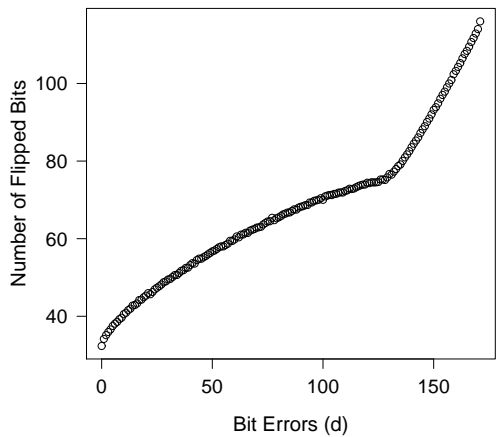}
\end{center}
\caption{Average number of flipped bits in the
channel as a function of the number $d$ of bit errors.}
\end{figure}

From this viewpoint, it will be more interesting to 
treat a channel of fixed number of flipped bits, instead
of binary symmetric channel. It is equivalent to 
the sampling of the positions of flipped bits under
the condition that the total number of flipped bits is given. 
An advantage of the proposed method is that we can
easily adapt it to this kind of modification.
In the present case, sampling from 
the channel of fixed number of flipped bits is
simply realized by the introduction of Metropolis move 
of swapping the positions of a flipped bit and a 
conserved bit. 

In Fig.4, a result by the proposed method in the case
of fixed number of flipped bits is shown.
The length of the original message and encoded message
is the same as the one 
of Fig.1--Fig.3, and 
the number of flipped
bits is set to $40$.  
The symbol \verb+(+$\bullet$\verb+)+ corresponds to 
the estimated probabilities by the proposed method. 
Preliminary runs with 9 iterations are performed 
for the tuning of the weight and the 10th run 
is used for the calculation of the results.
Total of these runs requires about $7.3 \times 10^8$ Viterbi
decoding. The symbol \verb|+| 
corresponds to the result by the naive method
with $1.1 \times 10^9$ Viterbi decoding. 
The largest value  $79$ of the bit errors
obtained by the proposed method seems to be the exact 
upper bound under the condition of Fig.4, because
it is stable against the increase of the weights for
$d>79$.     In the tail region shown in Fig.4, the probability 
is $\sim 10^{-20}$, which can never be estimated 
by the naive method. 

\begin{figure}[tb]
\vspace*{-0.3cm}
\begin{center}
\includegraphics[width=7cm]{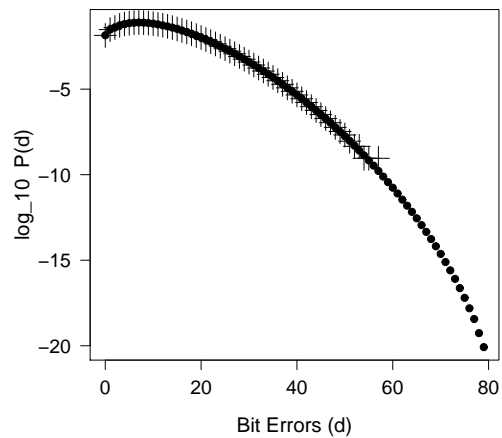}
\end{center}
\caption{Comparison of the estimates
of $P(d)$ calculated by the proposed method($\bullet$)
and the naive method(+). Fixed number of flipped bits.}
\end{figure}

In summary, we proposed an application of
the idea of rare event sampling to the estimation of the 
performance of error-correcting codes. 
It is shown that a method based on multicanonical
sampling of the pattern of noises gives an efficient
way for sampling of the tails of 
the distribution of bit errors with
given distributions of the input and noise.

A potential advantage of the proposed
approach is that we can explicitly
sample bit patterns of noises that
give severe ``damage'' to encoded
messages and cause
large bit errors. It can be useful for the understanding
of weak points of a given code. This idea of
``weak point sampling'' will be useful
for wide range of problems. Research in this
direction as well as applications to realistic
codes and channels are left for future studies.

This work is supported by the Grants-In-Aid for
Scientific Research (No. 17540348) from MEXT of Japan.


\begin{thebibliography}{99} 

\bibitem{Gbook}
R.~G.~Gallager:
{\it Information Theory and Reliable Communication},
(John Wiley \verb+&+ Sons, 1968).

\bibitem{Mbook}
D.~J.~C.~MacKay: {\it Information 
Theory, Inference, and Learning Algorithms}, 
(Cambridge University Press, 2003).

\bibitem{Nbook} H.~Nishimori: 
{\it Statistical Physics of Spin Glasses and
Information Processing: An Introduction},
(Oxford University Press, 2001).

\bibitem{BergC}
B.~A.~Berg and T.~Celik:
\textit{Phys.~Rev.~Lett.} \textbf{69} (1992) 2292.

\bibitem {BergN}
B.~A.~Berg and T.~Neuhaus: 
\textit{Phys.~Lett.~B} \textbf{267}, (1991) 249.

\bibitem {Lee}
J.~Lee:
\textit{Phys.~Rev.~Lett}, \textbf{71} (1993) 211.

\bibitem{iba}
Y.~Iba:
\textit{Int.~J.~Mod.~Phys.}, C~\textbf{12} (2001) 623.

\bibitem{cap1}
J.~D.~Lafferty and L.~A.~Wasserman: {\it Proc. of the 17th
Conference in Uncertainty 
in Artificial Intelligence} (2001) 293. 

\bibitem{cap2}
J. Dauwels:
{\it Proc. of the 26th Symposium on Information Theory 
in the BENELUX}, Brussels, Belgium (2005) 221.

\bibitem{Hartmann2002}
A.~K.~ Hartmann:
\textit{Phys. Rev. } E \textbf{65} (2002) 056102.  
 \bibitem{Keorner2006}
M.~K\"{o}rner,  H.~G.~Katzgraber, and A.~K.~Hartmann:
	 \textit{J. Stat. Mech.: Theory Exp.}  (2006) P04005.

\bibitem{Monthus2006}
C.~Monthus and T.~Garel:
\textit{Phys. Rev.} E \textbf{74} (2006) 051109.

\bibitem{H07}
K.~Hukushima and Y.~Iba:
\textit{J.~Phys.: Conference Series}, \textbf{95} (2008) 012005.

\bibitem{N07}
Y.~Matsuda, H.~Nishimori, and K.~Hukushima: arXiv:0712.4063.

\bibitem{xxx}
A zero-one asymmetry is
introduced in our implementation of the
Viterbi decoder by the choice of 0 and 1
when the weight is equal. It breaks 
the gauge invariance and considerably
affects the results quantitatively, 
while qualitative characteristics do not change.
This effect could be reduced by
averaging over the results with several values of 
$\mathbf{x}$ or treating
$\mathbf{x}$ as a stochastic variable.

\end{thebibliography}
\end{document}